\documentclass[12pt]{article}

\usepackage{scicite}
\usepackage{xcolor}

\usepackage{graphicx}
\usepackage{dcolumn}
\usepackage{bm}
\usepackage{relsize}
\usepackage{times}
\usepackage{siunitx}
\usepackage{float}
\usepackage{physics}
\usepackage{graphics}
\usepackage{tabularx}

\usepackage{longtable}
\usepackage{nicefrac}
\usepackage{amsmath}
\usepackage{amsfonts}
\usepackage{amssymb}
\usepackage{amsthm}
\usepackage{bm}
\usepackage{bbm}

\newenvironment{sciabstract}{%
\begin{quote} \bf}
{\end{quote}}

\newcommand{\hci}[3]{\textsuperscript{#1}#2\textsuperscript{#3}} 

\newcolumntype{.}{D{x}{}{-1}}
\newcolumntype{w}[1]{D{.}{.}{#1}}

\newcommand{\Za}{Z\alpha}

\newcommand{\vare}{\varepsilon}

\newcommand{\bfr}{{\bm r}}
\newcommand{\bfx}{{\bm x}}
\newcommand{\bfy}{{\bm y}}

\newcommand{\balpha}{\bm{\alpha}}

\newcommand{\bnabla}{\bm{\nabla}}

\newcounter{lastnote}
\newenvironment{scilastnote}{%
\setcounter{lastnote}{\value{enumiv}}%
\addtocounter{lastnote}{+1}%
\begin{list}%
{\arabic{lastnote}.}
{\setlength{\leftmargin}{.22in}}
{\setlength{\labelsep}{.5em}}}
{\end{list}}

\title{Testing inter-electronic interaction in lithium-like tin}

\author
{Jonathan Morgner,$^{\ast}$
Vladimir A. Yerokhin,
Charlotte M. König,
Fabian Hei{\ss}e,\\
Bingsheng Tu,$^\dag$
Tim Sailer,
Bastian Sikora,
Zoltán Harman,\\
José R. Crespo López-Urrutia,
Christoph H. Keitel,
Sven Sturm,
Klaus Blaum\\
\\
\normalsize{Max-Planck-Institut für Kernphysik, Heidelberg, Germany,}\\
\\
\normalsize{$^\ast$jonathan.morgner@mpi-hd.mpg.de}
\\
\normalsize{$^\dag$present address: Institute of Modern Physics, Fudan University, Shanghai, China}
}

\date{}

\begin{document} 


\baselineskip24pt


\maketitle

\begin{sciabstract}
Magnetic moments of bound-electron systems are a sensitive tool for testing fundamental interactions. $g$ factors of lithium-like ions have been rigorously studied in recent years, enabling insights into the relativistic inter-electronic effects. Here, we present the $g$-factor measurement of lithium-like tin, accurate to 0.5 parts per billion, as well as \textit{ab initio} theoretical calculations that include an advanced treatment of the inter-electronic interaction. We further improve the prediction by using the experimental result for the hydrogen-like tin $g$ factor, inferring from it the unknown higher-order QED effects. The observed agreement independently confirms the revised theory at a previously inaccessible high nuclear charge $Z$ of 50, where QED effects are significantly larger.
\end{sciabstract}

%
%
%
Quantum electrodynamics (QED) is the fundamental theory describing the electromagnetic interaction of charged particles by exchange of photons.
It includes non-classical effects such as vacuum polarization and the electron self-energy. Such effects have been studied in a wide variety of systems~\cite{beltrami_new_1986,beiersdorfer_measurement_2005,sturm_g_2011,kraft-bermuth_microcalorimeters_2018,fan:23,sailer_measurement_2022}], subjecting the QED framework to rigorous scrutiny.
Recently, the magnetic properties of lithium-like ions have raised interest both in experimental ~\cite{wagner_g_2013,kohler_isotope_2016,ullmann_high_2017,hannen_lifetimes_2019,micke_coherent_2020} and theoretical research~\cite{yerokhin:17:pra:segfact,yerokhin:20:gfact,yerokhin:21:gfact,kosheleva:22,zinenko:23}.
These ions are in many respects similar to hydrogen-like systems but are significantly richer because of the interaction not only between the electron and the nucleus but also among the electrons. Nonetheless, their configuration is still simple with a single $2s$ electron above a tightly-bound $1s^2$ shell. This makes the electron-electron interaction tractable within the QED theory, so that theoretical predictions can still challenge experiments to match their precision. Even more advantageous is to measure both the lithium-like and the hydrogen-like charge states of the same element, which allows the theory to be enhanced further by eliminating unknown contributions that are correlated in different electronic configurations. As a result, the comparison of theory and experiment for a combination of the lithium-like and the hydrogen-like charge states is capable of yielding enhanced tests of bound-state QED ~\cite{shabaev_g_2002} and possibly, in the future, an improved determination of the fine-structure constant~\cite{yerokhin:16:gfact:prl}, as well as placing stringent bounds on the coupling constants of beyond Standard Model-interactions~\cite{debierre_fifth-force_2020}.
\newline

{\em Ab initio} QED calculations of the $g$ factor of lithium-like ions are very cumbersome and proved to be problematic in the past.
In particular, several advanced QED calculations~\cite{glazov:19,yerokhin:20:gfact,yerokhin:21:gfact} reported
up to 5 standard-deviations from the experimental data on lithium-like
silicon ($Z=14$) and calcium ($Z=20$).
Recently, a careful re-analysis of the interelectronic QED effects in Ref.~\cite{kosheleva:22} found agreement with the calcium experiment
and a deviation of only 1.5~$\sigma$ for silicon. 
However, it remains unclear if the problem was fully resolved, until a test of theory with a new measurement, ideally on a lithium-like ion with a higher $Z$, is accomplished.
\newline

\begin{figure}[H]
            \centering
            \includegraphics[width=90mm]{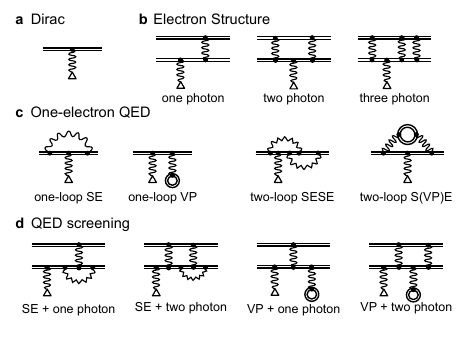}
            \caption{
                Main Feynman diagrams describing the $g$ factor of a few-electron ion.
                The double lines denote the bound electrons, the wavy lines represent the virtual photon exchange, and
                the wavy lines terminated by a triangle indicate the interaction with the external magnetic field.
            }
            \label{fig:diagrams}
            \hfill
\end{figure}

Here we report on a high-precision $g$-factor measurement of the much heavier lithium-like system \hci{118}{Sn}{47+}, accompanied with its {\em ab initio} QED calculation.
The nuclear charge of tin with $Z=50$ and, correspondingly, the nuclear electric field, are much higher than in the previous $g$-factor measurements, testing the QED theory of bound states in a very different regime.
\newline\newline
\textbf{QED theory of the lithium-like tin \textit{g} factor \newline}
In {\em ab initio} bound-state QED theory, the $g$ factor of the lithium-like ion arises in zeroth order from the interaction of the spin of the $2s$ valence electron with the external magnetic field.
This explains why it is numerically close to that of the free electron $g_e$, which is known with great accuracy both experimentally and theoretically \cite{fan:23,aoyama:18}.
Therefore, we focus here on the so-called binding effects, which result from the interaction with the nucleus and other electrons in the atom.
The largest binding contribution follows from the Dirac equation and was analytically described for the point nucleus by G.~Breit in 1928 \cite{breit_magnetic_1928},
\begin{equation}
  \label{eq:gd}
  \delta g_{\mathrm{D}}(\mbox{\rm pnt}) = \frac{2}{3}\left( \sqrt{2\left(1+\sqrt{1-(Z\alpha)^2}\right)} -2 \right)\,,
\end{equation}
where $Z$ is the atomic number and $\alpha$ is the fine-structure constant.
In addition, the $g$ factor receives numerous smaller corrections from QED, the electron-electron interaction and finite nuclear size and mass.
The contributions are summarized in Table~\ref{tab:gtheo}, and the main Feynman diagrams are shown in Fig.~\ref{fig:diagrams}.
\newline

For convenience we consider two classes of binding effects separately: one-electron contributions induced exclusively by the valence $2s$ electron and the many-electron corrections caused by the interaction with the other bound electrons.
The dominant one-electron QED corrections induced by the one-loop electron self-energy and vacuum polarization have been extensively studied and can be calculated very accurately~\cite{yerokhin:04,yerokhin:08:prl,yerokhin:17:pra:segfact}.
The effects of the finite nuclear size and mass are also well under control~\cite{shabaev:02:recprl,shabaev:17:prl}. 
However, the two-loop QED corrections are far more challenging.
Despite a lot of attention in recent years~\cite{pachucki:05:gfact,yerokhin:13:twoloopg,czarnecki:16,czarnecki:18,czarnecki:20,sikora:20}, a large part of the two-loop QED effects is calculated only within the $Z\alpha$ expansion.
For tin with $Z\alpha \approx 50/137 \approx 0.36$, the expansion is poorly converging and the uncertainty on the two-loop QED contribution is therefore rather large.
However, in a previous investigation~\cite{morgner_stringent_2023} the missing two-loop effects for the $1s$ state of hydrogen-like tin were isolated experimentally.
This allows us to use the rescaled experimental two-loop QED correction to reduce the uncertainty of the theoretical prediction for the lithium-like ion.
\newline

What makes the calculations of the $g$ factor of a lithium-like ion truly challenging is the presence of multiple electrons.
An accurate theory of the interelectronic effects in the atomic $g$ factor requires a systematic treatment of the non-local electron-electron interaction, which is possible only within QED.
The QED perturbation expansion is formulated in terms of Feynman diagrams containing exchange of one, two, three, etc.~virtual photons between the electrons.
The resulting multitude of diagrams is divided into two classes: those without radiative loops, called electron-structure corrections, and those with self-energy or vacuum-polarization loops, called QED-screening corrections (see Fig.~\ref{fig:diagrams}).
Rigorous calculations of the two-photon electron-structure and one-photon QED screening diagrams have recently become possible and were carried out without any expansion in the parameter $Z\alpha$ 
\cite{volotka:14,glazov:19,yerokhin:20:gfact,yerokhin:21:gfact,kosheleva:22}.
In the present work, we refined that approach, leading to improved accuracy. 
In particular, the numerical precision for the self-energy screening correction for $Z = 14$ and $20$ \cite{kosheleva:22} has been improved by a factor of 10, which removed one of the dominant errors in the theoretical predictions (See Supplementary Material).
This improvement was made possible by using a highly accurate Green-function representation of electron propagators and including in it 2.5 times more partial waves than in previous computation
\newline

With the leading electron structure and QED screening effects accurately calculated, the source of theoretical uncertainty shifts to higher-order interelectronic effects, which currently cannot be calculated rigorously to all orders in $Z\alpha$.
The magnitude of these effects has previously been underestimated.
These effects were responsible for the deviations of theory and experiment for silicon and calcium~\cite{glazov:19,yerokhin:20:gfact,yerokhin:21:gfact}, as argued in Ref.~\cite{kosheleva:22} where they applied recursive perturbation theory~\cite{glazov:17} and obtained approximate results for the corresponding corrections for silicon and calcium. 
In the present work, we employ the non-relativistic QED (NRQED) approach to evaluate them to the leading order in $Z\alpha$ with a very high numerical accuracy (supplementary material).

The crucial point now is to estimate the theoretical uncertainty stemming from the omitted higher-order terms in $Z\alpha$ interelectronic effects.
We do this by varying the zeroth-order approximation, namely by including the so-called screening potential (approximately describing the interaction with core electrons) in the zeroth-order Dirac equation.
Our uncertainty estimates are based on the spread of results obtained with different screening potentials (supplementary material).
\newline

Adding all the binding effects and the well-established value of the free-electron $g$ factor \cite{fan:23,aoyama:18}, we obtain the total theoretical prediction for the $g$ factor of $^{118}$Sn$^{47+}$ as
\begin{equation} \label{eq:theo:1}
g_\mathrm{theo} = 1.980\,354\,769\,(35)\,.
\end{equation}
As can be seen from Table~\ref{tab:gtheo}, in our calculations we were able to compute the interelectronic effects with the absolute error of $8\times 10^{-9}$, so that
the dominant theoretical uncertainty
now comes from the one-electron two-loop QED effects. 
This is in sharp contrast to previous measurements
of lighter lithium-like ions \cite{wagner_g_2013,glazov:19}, where the theoretical predictions were limited by
the electron-structure effects.
\newline

The {\em ab initio} prediction (\ref{eq:theo:1}) can be further improved if we use the two-loop QED contribution extracted from the recent measurement of the $g$ factor in hydrogen-like tin~\cite{morgner_stringent_2023} and scale it to the lithium-like electron case. 
This procedure is similar to the specific difference used to reduce uncertainties from  nuclear-size contributions that behave similar in different charge states of the same element~\cite{shabaev_g-factor_2006}.
Performing the scaling conversion (supplementary material),
we obtain an ``experimentally enhanced'' theoretical prediction
\begin{equation}\label{eq:theo:2}
g_\mathrm{theo}(\mathrm{enh}) =  1.980\,354\,796\,(12)\,,
\end{equation}
with three times smaller uncertainty than the purely theoretical value (Eq.~\ref{eq:theo:1}).
Because the problematic higher-order two-loop QED effects are about 8 times smaller for the lithium-like ions than for the corresponding hydrogen-like configuration, we achieve in total a relative prediction accuracy of $6\times10^{-9}$, a 25-fold improvement over the hydrogen-like case.
\newline\newline
\textbf{Experimental measurement of the $g$-factor}\newline
For the high-precision measurement, the ions are produced externally in an electron beam ion trap, where electrons are stripped from the nucleus by electron impact ionization. The highly charged ions are transported into the \textsc{Alphatrap} Penning-trap apparatus through a room-temperature beamline.
In this case hydrogen-like tin was captured, and subsequent charge exchange resulted in double electron-capture, resulting in lithium-like tin.
In the Penning Trap a the strong magnetic field $B_0\approx\SI{4}{\tesla}$ in the \textsc{Alphatrap} apparatus confines a single ion to a cyclotron orbit allowing us to perform non-destructive Penning-trap spectroscopy~\cite{sturm_alphatrap_2019}.
The gyromagnetic ratio, or $g$ factor of the spin~$\nicefrac 1 2$ particle is determined by measuring its Larmor precession frequency $\nu_\mathrm{L}$ together with $B_0$; the two quantities are connected by the relation
\begin{equation}
    \nu_\mathrm{L} = \frac{g}{4 \pi}\frac e {m_e} B_0,
\end{equation}
where $\frac{e}{m_e}$ is the electron charge-to-mass ratio.
$B_0$ is measured via the cyclotron frequency $\nu_c = \frac 1 {2 \pi }\frac q {M} B_0$, where $\frac{q}{M}$ is the charge-to-mass ratio of the ion.
Combined, this results in:
\begin{equation}
    g = 2~\frac{\nu_\mathrm{L}}{\nu_\mathrm{c}}\frac q e \frac{m_e}{M}.\label{eq:g}
\end{equation}
Therefore, to determine the $g$ factor we measure the ratio $\frac{\nu_\mathrm{L}}{\nu_c}=\Gamma_0$ and combine it with the literature values for the other parameters~\cite{tiesinga_codata_2021,morgner_stringent_2023}.
\newline

\begin{figure}[H]
    \centering
    \includegraphics[width=\textwidth]{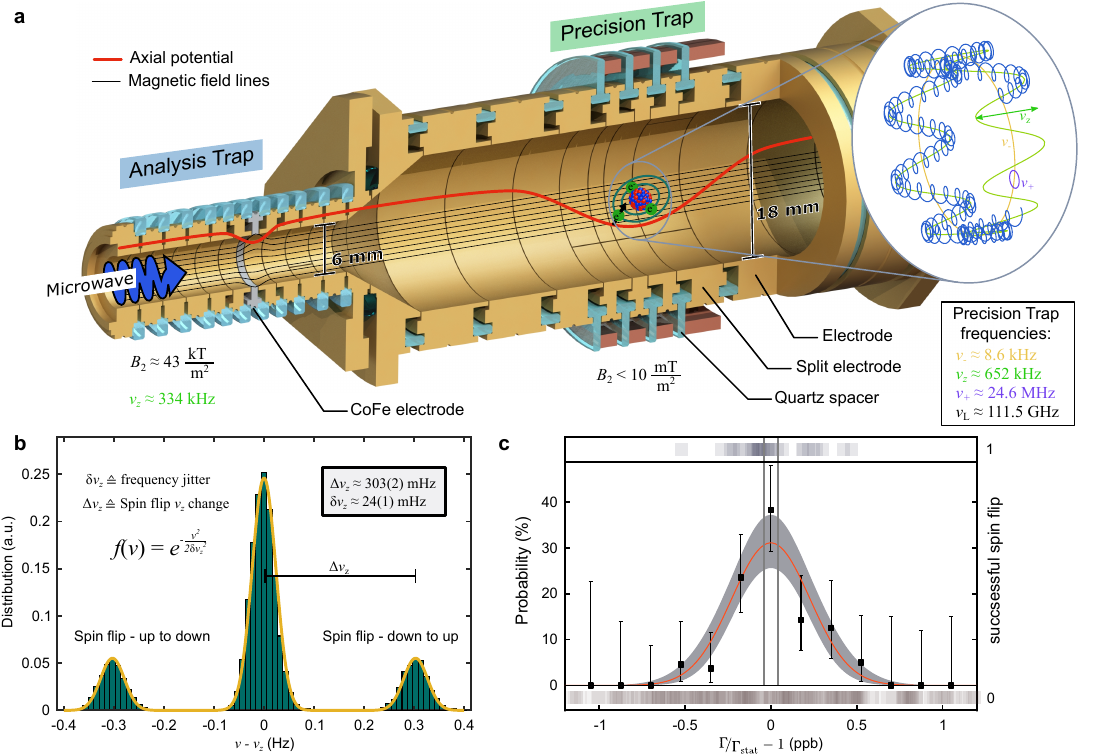}
    \caption{
        \textbf{Schematic and results of the experiment.}
        \textbf{a} The precision measurement takes place in two separate traps, between which the ion is shuttled by adiabatic transport.
        In the Precision Trap, the electric field is extremely harmonic and the magnetic field is as homogeneous as possible, which are optimal conditions for precision spectroscopy.
        In the Analysis Trap the center electrode is made of a ferromagnetic material that produces a large magnetic field inhomogeneity, allowing non-destructive detection of the spin-state~\cite{dehmelt_continuous_1986}.
        \textbf{b} The frequency data taken in the Analysis Trap during the $g$-factor measurement.
        As the likelihood to drive a spin flip (see main text) is only 50\%, half of the time no frequency change is observed.
        When a spin flip occurs, the frequency is 300~mHz larger or smaller, corresponding to the distributions at +300~mHz and -300 mHz.
        Based on direction of the change, the current (and prior) spin orientation is determined.
        \textbf{c} Fit to the recorded spin-flip data.
        The resonance consists of 330 individual measurement points, shown at the top and bottom for successful and unsuccessful attempts, respectively.
        The black scattered dots are a binned set of the data, shown for better visualization.
        The red line is the fitted curve, and the grey band shows its 1-sigma confidence interval.
        The grey vertical lines show the 1-sigma confidence intervals of $\Gamma_\text{stat}$.
    }
    \label{fig:trap}
\end{figure}

In addition to the magnetic radial confinement, an electrostatic field traps the ion in the axial direction (see Fig.~\ref{fig:trap}~\textbf{b}).
As a result of the optimized electrode geometry, the ion motion can be described by a close-to-perfect harmonic oscillator within a large region of the trap.
Along the magnetic field lines, the ion oscillates with an axial frequency $\nu_z$.
The $\textbf{E}\cross\textbf{B}$ drift caused by the applied electric field slightly modifies the frequency of the cyclotron motion, and an additional magnetron motion with frequency $\nu_-$ appears.
The resulting particle trajectory is shown in the inset of Fig.~\ref{fig:trap}~\textbf{a}.
The three eigenmodes - modified cyclotron, axial and magnetron - are related to the free-space cyclotron frequency by the invariance theorem which shows that errors caused by misalignment of the electric and magnetic fields are cancelled, and the motions are connected via the equation $\nu_\mathrm{c}^2 = \nu_+^2+\nu_z^2+\nu_-^2$~\cite{brown_geonium_1986}.
The axial frequency is measured via the femtoampere image current induced in the surrounding electrodes.
It is transformed into a measurable voltage by a superconducting tank circuit which interacts with the ions when in resonance, resulting in energy exchange and consequently thermalization of the axial motion to the temperature of the detector at $\SI{5.7(2)}{\kelvin}$.
In this case, the Fourier spectrum of the detector signal shows a distinct dip at the mode frequency, which can then be determined from fitting.  
The cyclotron and magnetron frequencies are measured with the same detector using radio-frequency quadrupole fields, which couple the modes.
As the frequency hierarchy in a Penning trap is typically $\nu_- \ll \nu_z \ll \nu_+$, $\nu_\mathrm{c}$ is mostly dependent on the frequency $\nu_+$ of the modified cyclotron motion.
Frequencies of the $^{118}\mathrm{Sn}^{47+}$ ion are shown in Fig.~\ref{fig:trap}~\textbf{a}.
To determine the ratio of Larmor to cyclotron frequency $\Gamma_0$, we measure $\nu_+$ and simultaneously irradiate a microwave.
When the ratio of the irradiated microwave frequency $\nu_\text{MW}$ and the measured $\nu_\text c$ are in resonance with $\Gamma_0$, the transition is driven, and the electron spin can change its orientation.
To detect such a spin flip non-destructively, we use the continuous Stern-Gerlach effect~\cite{dehmelt_continuous_1986}.
In the nearby Analysis Trap, a large magnetic bottle field exerts a force on the magnetic moment of the ion.
Depending on the spin orientation with respect to the magnetic field, the ion becomes 'low-field' or 'high-field-seeking'.
Thus a spin flip results in a shift of the axial frequency (see Fig.~\ref{fig:trap}~\textbf{b}):
    \begin{equation}\label{eq:deltawz}
        \Delta \nu_z \approx \frac {B_2\hbar q}{4 \pi^2 M^2 \nu_z}\Gamma_0 \Delta m_s.
    \end{equation}
The magnetic bottle strength, $B_2 \approx 43\,\frac{\mathrm{kT}}{\mathrm{m^2}}$, is the second order magnetic field coefficient $B(z) = B_0 + B_2z^2 + ...$ in the Analysis Trap.
$\nu_z \approx 334\,\mathrm{kHz}$ is the axial frequency in the trap, and the change of the spin magnetic quantum number $\Delta m_s$ is $\pm 1$ is depending on the initial spin orientation.
Fig.~\ref{fig:trap} \textbf{a} highlights the magnetic field lines in the Analysis Trap, where a direct high-precision measurement of $\Gamma_0$ is not possible because of the magnetic field inhomogeneity.
Therefore, we measure $\Gamma_0$ in the Precision Trap, employing the double-trap technique~\cite{haffner_double_2003}.
The magnetic field in this trap is much more homogeneous.
Combined with an excellent harmonicity of the electrostatic trapping field, systematic effects in this trap are quite small and under better control (supplementary material).
The double-trap setup for the measurement is shown in Fig.~\ref{fig:trap}~\textbf{a}, the microwaves are injected through a millimeter waveguide into the setup, irradiating all traps with the electromagnetic wave.
\newline

The measurement sequence of $\Gamma_0$ follows the one presented in Ref.~\cite{morgner_stringent_2023}.
The particle is continuously shuttled between the two traps.
First we determine its spin state in the Analysis Trap.
To achieve this with high fidelity, we set the axial oscillation frequency in resonance with the detector circuit by changing the voltage applied to the electrodes.
We then measure the motional frequency, followed by a high-power microwave to try to change the spin state.
This drive quickly loses coherence and has a roughly $50\,\%$ chance of flipping the spin in the Analysis Trap.
The axial frequency is measured again to detect a possible spin flip, observable as a jump in frequency after the microwave irradiation.
We use a phase-sensitive technique to measure the axial frequency jump as described in Ref.~\cite{stahl_phase-sensitive_2005}, allowing the spin state to be read quickly.
Fig.~\ref{fig:trap}~\textbf b shows the frequency readings of the spin-state detection during the measurement run with a measured shot-to-shot stability of about $24(1)\,\text{mHz}$.
A spin flip shifts the frequency by about $300\,\text{mHz}$ (Eq.~(\ref{eq:deltawz})).
This, together with averaging over four phase measurements results in extremely low error rates in state detection.
After state determination, the sequence continues in the Precision Trap where the motional eigenfrequencies are measured.
During the $\nu_+$ measurement the microwave field at frequency $\nu_\mathrm{MW}$ is applied, chosen with a random offset to the expected Larmor frequency.
We then check in the Analysis Trap whether the spin state was flipped by this.
Repeating the cycle, which takes on average 20 minutes, at different ratios $\Gamma$ results in a probability histogram, shown in Fig.~\ref{fig:trap}~\textbf c.
A total of 330 $\Gamma$ ratios were probed, and 31 resulted in successful spin flips.
Using a maximum-likelihood analysis we determine the center to be $\Gamma_\mathrm{stat} = 4526.894\,266\,68(20)$.
Correcting this for systematic effects as shown in Table~\ref{tab:ExpData}, we extract a measured value for the frequency ratio:
\begin{equation}
    \Gamma_0 = 4526.894\,265\,905(191)_\mathrm{stat}(122)_\mathrm{sys}.
\end{equation}
The treatment of systematic effects is summarized in the supplementary material. This follows the full discussion as given in Ref.~\cite{morgner_stringent_2023}, which uses very similar trap parameters, resulting in near identical systematic shifts and effects.
Using Eq.~(\ref{eq:g}), the literature value of the electron mass~\cite{tiesinga_codata_2021} and the ion mass reported in \cite{morgner_stringent_2023}, corrected by the two additional electrons and binding energies~\cite{kramida_nist_2021}, we determine the $g$ factor of the lithium-like tin ion as:
\begin{equation}\label{eq:exp}
g_\mathrm{exp} = 1.980\,354\,799\,750(84)_\mathrm{stat}(54)_\mathrm{sys}(944)_\mathrm{ext}.   
\end{equation}
The brackets are the 1-sigma confidence intervals of the statistical, systematic and external uncertainties, respectively.
The external uncertainties are dominated by the ion mass uncertainty which currently limits the experimental accuracy.
It is also feasible to improve the precision of the mass value by more than an order of magnitude and consequently enhance the precision of the measured $g$ factor~\cite{rischka_mass-difference_2020,kromer_high-precision_2022,heise_high-precision_2023}.\newline\newline
\textbf{Discussion and outlook }\newline

Experimental data was blinded at the time of the theory evaluation.
The observed agreement between experiment and theory (Eq.~\ref{eq:exp} and \ref{eq:theo:1},\ref{eq:theo:2}) is an important step forward as it consolidates the improvement of the most recent $g$-factor calculations over previous theoretical studies, which repeatedly reported results in discrepancy with measured values~\cite{glazov:19,yerokhin:20:gfact,yerokhin:21:gfact,kosheleva:22}.
Our results provide a crucial test of sophisticated many-body QED calculations performed for a much higher nuclear charge $Z$ than previously, thus probing the non-trivial bound-state QED effects scaling as $Z^4$ to a significantly greater extent.
Furthermore, the agreement of the enhanced theoretical value  (Eq.~\ref{eq:theo:2}) with the  experiment implicitly checks the consistency of the experiments on hydrogen-like and lithium-like tin and their theory.\newline
The improved precision of the theoretical prediction allows us to clearly resolve the electron-structure and QED screening effects and to test the advanced calculations in the so far unexplored regime of medium-to-high $Z$ ions.
In the future, measurements of heavier lithium-like systems such as \hci{208}{Pb}{79+} and the expected progress in two-loop QED calculations will explore with even higher precision the regime of the strong electric fields reachable with highly charged ions.
The advanced theoretical methods developed in this work for the description of interelectronic QED effects can be applied to calculations of $g$-factors of more complex ions (boron-like, carbon-like~\cite{arapoglou_g_2019,ullmann_high_2017}), parity-nonconserving transition amplitudes in neutral atoms and other effects~\cite{safronova_search_2018}.

\begin{table}[H]
    \centering
    \footnotesize
    \caption{
        Binding corrections. Also shown are total theoretical values of the $g$ factor of $^{118}$Sn$^{47+}$,
        with and without the ``enhancement'' by the experimentally measured two-loop effects (see text).
        }
    \begin{tabularx}{65mm}{l r}
         \hline\hline
         Effect & Value ($\times 10^{-6}$)\phantom{(xx)}\\\hline
         Dirac, point nucl.   & $-$23\,181.721\phantom{(xx)}\\
         \,\,\,-\,\,\, ,  FNS &          2.040\,(3)\phantom{x}\\
         Electron structure   &     1\,192.179\,(7)\phantom{x}\\
         One-loop QED         &         23.977\,(1)\phantom{x}\\
         Two-loop QED         &       $-$0.107\,(33)  \\
         \,\,\,-\,\,\, , enhanced &   $-$0.080\,(6)\phantom{x}\\
         QED screening        &       $-$1.076\,(8)\phantom{x}\\
         Nuclear recoil       &          0.172\,(1)\phantom{x}\\
         $g_e$ \cite{fan:23,aoyama:18}  & 2\,002\,319.304\,361\phantom{x\!\!}\\
         \hline
         $g$, theory          & 1\,980\,354.769\,(35)\\
    \,\,\,-\,\,\, , enhanced  & 1\,980\,354.797\,(12)\\
         \hline\hline
    \end{tabularx}
    \label{tab:gtheo}
\end{table}

\begin{table}[H]
    \centering
    \footnotesize
    \caption{
        \textbf{Error budget of $\Gamma_0$ and $g$.}
        Following Ref.~\cite{morgner_stringent_2023}, additional shifts and uncertainties are smaller than \SI{1}{}~ppt, and thus insignificant here.
    }
    \begin{tabularx}{120mm}{l r r}
        \textbf{Parameter} & \textbf{Rel. shift (ppt)}  & \textbf{Uncertainty (ppt)} \\ \hline\hline
        $\Gamma = \nu_L/\nu_c$ error budget: & &\\ 
        \hspace{0.5cm}$\nu_-$ measurement & -\phantom{.0}\ & 4.3\\
        \hspace{0.5cm}Relativistic shift~\cite{ketter_classical_2014} & 21.8 & 4.4\\
        \hspace{0.5cm}Image-charge shift~\cite{schuh_image_2019} & 148\phantom{.0}\ & 7.5 \\ 
        \hspace{0.5cm}$\nu_z$ line shape & -\phantom{.0}\ & 22\phantom{.0}\\ 
        \hspace{0.5cm}statistical uncertainty & -\phantom{.0}\ & 42\phantom{.0}\\ \hline
        $g$-factor error budget: & &\\ 
        \hspace{0.5cm}Total $\Gamma_0$ uncertainty & &44\phantom{.0}\\ 
        \hspace{0.5cm}Electron mass~\cite{sturm_high-precision_2014,tiesinga_codata_2021} &  & 29\phantom{.0} \\ 
        \hspace{0.5cm}\hci{118}{Sn}{47+} mass~\cite{morgner_stringent_2023,kramida_nist_2021}  & & 475\phantom{.0}\\   \hline \hline 
    \end{tabularx}
    \label{tab:ExpData}
\end{table}


\begin{scilastnote}
\item {Acknowledgements}
 This work was supported by the Max Planck Society (MPG), the International Max Planck Research School for Quantum Dynamics in Physics, Chemistry and Biology (IMPRS-QD), the German Research Foundation (DFG) Collaborative Research Centre SFB 1225 (ISOQUANT) and the Max Planck PTB RIKEN Center for Time, Constants, and Fundamental Symmetries.
 This project has received funding from the European Research Council (ERC) under the European Union’s Horizon 2020 research and innovation programme under grant agreement number 832848 FunI.
 This work comprises parts of the PhD thesis work of J.M. to be submitted to Heidelberg University, Germany.
\end{scilastnote}

\newpage\null\thispagestyle{empty}\newpage

\section{Supplementary Material}

In the zeroth-order approximation, the $g$ factor of a lithium-like ion
arises through the interaction of the $2s$ valence electron with the external magnetic field.
It can be represented as an expectation value of the effective magnetic interaction
$V_g$ with the Dirac wave functions of the valence electron $v$,
\begin{align}\label{eq:01}
g_\mathrm{D} = \bra{v}  V_g(\bfr)\ket{v} \equiv \bra{v} \frac{1}{\mu_v}\, (\bfr \times\balpha)_z \ket{v}  \,,
\end{align}
where $\bm{r}$ is the position vector,
$\balpha$ is the vector of Dirac matrices, and $\mu_v$ is the angular momentum projection of
the valence electron.
For a point-like nucleus, the matrix element can be evaluated analytically, with the result
\begin{equation}
  \label{eq:4}
  g_{\mathrm{D}}(\mbox{\rm pnt}) = \frac{\kappa_v}{2j_v(j_v+1)}\Big( 2\kappa_v \frac{\varepsilon_v}{m} - 1 \Big)\,,
\end{equation}
where $\kappa_v$ is the relativistic angular-momentum quantum number of the state $v$, $j_v = |\kappa_v|-1/2$,
$\varepsilon_{v}$ is the Dirac energy of the valence state, and $m$ is the electron
rest mass.
The nuclear size correction to the point-nucleus Dirac value is
calculated numerically, by taking the difference of
the matrix element (\ref{eq:01}) evaluated numerically for an extended nuclear charge distribution and the
point-nucleus result of Eq.~(\ref{eq:4}).

%
\subsection{One-electron QED effects}

The leading QED effects are the one-loop self energy (SE) and vacuum polarization (VP).
The self-energy correction to the bound-electron $g$ factor
is represented \cite{yerokhin:04}
as a sum of the irreducible (ir) and the vertex$+$reducible (vr) parts,
\begin{align}\label{se4}
g_{\rm se} = g_{\rm ir} + g_{\rm vr}\,.
\end{align}
The irreducible part is expressed in terms of a non-diagonal matrix element of the
renormalized one-loop
self-energy operator $\Sigma_R$,
\begin{equation}\label{se5}
g_{\rm ir} = 2\,\bra{v}\Sigma_R(\vare_v) \ket{\delta v}\,,
\end{equation}
where $\Sigma_R(\vare) = \Sigma(\vare) - \gamma^0\delta m$,
$\delta m$ is the one-loop mass counter term, $\gamma^0 = \beta$ is a
Dirac matrix in the usual notation, and $\ket{\delta v}$ is the magnetically
perturbed wave function of the valence state,
\begin{equation}
  \label{eq:13}
  \ket{\delta{v}}
  = \sum_{n \neq v}
  \frac{ \ket{n} \bra{n}V_{g}\ket{v}}{\varepsilon_v - \varepsilon_n} \,.
\end{equation}
The one-loop self-energy operator is defined as
\begin{eqnarray} \label{se6}
 \Sigma(\vare,\bfr_1,\bfr_2) &=&
2\,i\alpha\, \int_{-\infty}^{\infty} d\omega\,
      \alpha_{\mu}\,
         G(\vare-\omega,\bfr_1,\bfr_2)\, \alpha_{\nu}\,
    D^{\mu\nu}(\omega,\bfr_{12})\,,
\end{eqnarray}
where $G$ denotes the Dirac Coulomb Green's function $G(\vare) = [\vare-{\cal H}(1-i 0)]^{-1}$, $\cal
H$ is the Dirac Coulomb Hamiltonian, and $D^{\mu\nu}$ is the photon propagator.

The vertex$+$reducible part is given by
\begin{align}\label{se8}
g_{\rm vr} = &\ \frac{i}{2\pi} \int_{-\infty}^{\infty} d\omega\, \sum_{n_1n_2} \biggl[
  \frac{\bra{n_1}V_g \ket{n_2}\bra{vn_2} I(\omega)\ket{n_1v}}{(\Delta_{vn_1}-\omega)(\Delta_{vn_2}-\omega)}
   -
  \delta_{n_1n_2}\,\frac{\bra{v} V_g \ket{v} \bra{vn_1} I(\omega) \ket{n_1v} }{(\Delta_{vn_1}-\omega)^2}
  \biggr]\,,
\end{align}
where $\Delta_{vn} = \vare_v-\vare_n(1-i0)$, summations over $n_{1,2}$ are performed over the
complete spectrum of the Dirac equation, and
$I(\omega)$ is the operator of the electron-electron interaction,
\begin{equation}\label{a1}
  I(\omega,\bfr_{1},\bfr_{2}) = e^2\, \alpha_{1}^{\mu} \alpha_{2}^{\nu}\, D_{\mu\nu}(\omega,\bfr_{1},\bfr_2)\,,
\end{equation}
with $\alpha^{\mu}_{a} = (1,\balpha_{a})$ being the four-vector of Dirac matrices acting on
$\bfr_a$.

The vacuum-polarization correction is given by the sum of the electric-loop and magnetic-loop
contributions \cite{beier:00:rep},
\begin{align}
g_{\rm vp} = 2\,\bra{v} U_{\rm el} \ket{\delta v} +
\bra{v} V_g(\bfr)\,\Pi(r) \ket{v}\,,
\end{align}
where $U_{\rm el}$ is the electric-loop vacuum-polarization potential defined by
\begin{align}
U_{\rm el}(x) = \frac{i\alpha}{2\pi}\,\int d\bfy\,\frac1{|\bfx-\bfy|}\,
 \int_{-\infty}^{\infty}d\omega\, {\rm Tr}\, G(\omega,\bfy,\bfy)\,.
\end{align}
To the leading non-vanishing order in $\Za$,
the polarization function $\Pi(r) $ is given by  \cite{lee:05}
\begin{align}
\Pi(r) = \frac{\alpha}{\pi}\,(\Za)^2\,\frac{4}{r}\,\int_0^{\infty}dq\,q\,F(q)\,
j_1(qr)\,,
\end{align}
where $j_l(z)$ is the spherical Bessel function and the momentum-distribution function
$F(q)$ was calculated in \cite{lee:07}.

The one-electron self-energy correction was calculated here using the numerical
approach developed in Refs.~\cite{yerokhin:17:pra:segfact,yerokhin:16:gfact:pra}.
The electric-loop part of the one-electron vacuum polarization is relatively straightforward
to calculate (see, e.g., Ref.~\cite{cakir:20}). The magnetic-loop vacuum-polarization
correction was computed as described in Ref.~\cite{cakir:20}, using the analytical results for the momentum-distribution function $F(q)$
derived in Ref.~\cite{lee:05}.

%
\subsection{QED screening effects}

The presence of core electrons in lithium-like ions modifies the one-electron self-energy and vacuum-polarization
corrections considered in the previous section, which is known as the {\em QED screening} effect. It
can be accounted for
by the perturbation expansion in the electron-electron interaction. The self-energy and vacuum-polarization
screening corrections
(``sescr'' and ``vpscr'') are represented as
\begin{align} \label{se9}
g_{\rm sescr} = g^{(0)}_{\rm sescr} + g^{(1)}_{\rm sescr} + g^{(2)}_{\rm sescr} + \ldots \,,
 \\
g_{\rm vpscr} = g^{(0)}_{\rm vpscr} + g^{(1)}_{\rm vpscr} + g^{(2)}_{\rm vpscr} + \ldots \,,
\end{align}
where the upper index denotes the order in the residual electron-electron interaction,
i.e., the number of virtual photons exchanged between the electrons.
The zeroth-order terms $g^{(0)}$ appear when a screening potential is included
into the Dirac equation in the zeroth-order approximation. In this case, $g^{(0)}_{\rm sescr}$
($g^{(0)}_{\rm vpscr}$)
are given by
the difference of the one-electron corrections $g_{\rm se}$ ($g_{\rm vp}$)
calculated for the actual starting potential and the Coulomb potential.

Feynman diagrams with one photon exchange between
the electrons induce
the one-photon screening correction $g^{(1)}_{\rm sescr}$. General expressions for this correction were derived in Ref.~\cite{glazov:10}
and, in a slightly different formulation, in Ref.~\cite{yerokhin:20:gfact}, where the corresponding formulas can be reviewed.
Numerical calculations of the corresponding corrections
are rather difficult. The first one was reported in
Ref.~\cite{glazov:10} for lithium-like lead and uranium ions for the Coulomb starting potential. It used the $B$-spline finite-basis-set representation of the Dirac-Coulomb electron
propagators \cite{johnson:88,shabaev:04:DKB}. In the subsequent investigations
\cite{volotka:14,glazov:19,kosheleva:22}, this approach was extended to the case of the general screening
potential and applied to the silicon and calcium ions. An alternative numerical approach
based on the Dirac-Coulomb Green's function representation of the electron propagators was developed
in Ref.~\cite{yerokhin:20:gfact} for the Coulomb starting potential. In the present work
we extend this approach to the case of a general screening potential. Numerical techniques
for those computations with the Green's function are described in Ref.~\cite{yerokhin:20:green}.

The second-order self-energy screening contribution $g^{(2)}_{\rm sescr}$ contains
an exchange of two photons between the valence and core
electrons.
In the present work, we calculate this contribution to the lowest order in the parameter
$\Za$.
For the Coulomb starting potential, a similar calculation was performed in
Ref.~\cite{yerokhin:17:gfact} within the non-relativistic QED (NRQED) approach.
Kosheleva {\em et al.} \cite{kosheleva:22}
employed a different approach which is based on  recursive perturbation theory \cite{glazov:17} and
uses the effective Hamiltonian operators derived by Hegstrom \cite{hegstrom:73}.
In this work, we also use the Hegstrom operators but perform calculations within the
standard many-body perturbation theory.

\begin{table*}
\centering
\caption{The 
self-energy and vacuum-polarization
screening corrections for the $g$-factor of lithium-like tin, for different starting potentials,
in units $10^{-6}$. CH, KS$_0$, DH$_0$, DS$_0$ denote different choices of the screening potential in the zeroth-order approximation (the core-Hartree, Kohn-Sham, Dirac-Hartree, Dirac-Slater potentials, respectively).\\
\label{tab:scr}
}
\begin{tabular}{lw{2.7}w{2.7}w{2.7}w{2.7}w{2.7}}
     \multicolumn{1}{c}{Term} &
         \multicolumn{1}{c}{Coulomb} &
             \multicolumn{1}{c}{CH} &
                 \multicolumn{1}{c}{KS$_0$} &
                     \multicolumn{1}{c}{DH$_0$} &
                         \multicolumn{1}{c}{DS$_0$} \\
                         \hline\\[-5pt]
 $g^{(0)}_{{\rm sescr}}$           &               & -1.1768\,(3)   & -1.0199\,(3)   & -1.1782\,(3) &   -0.9418\,(3)  \\
 $g^{(1)}_{{\rm sescr}}$           & -1.2546\,(5)  & -0.1588\,(1)   & -0.3200\,(4)   & -0.1577\,(8) &   -0.4002\,(2)  \\
 $g^{(2)}_{{\rm sescr}}$           &  -0.059\,(59)  & -0.0001\,(20)  &  0.0022\,(22)  & -0.0033\,(33)&    0.0055\,(55) \\
 $g_{{\rm sescr}}$                 &  -1.314\,(59)  & -1.3357\,(20)  & -1.3377\,(23)  & -1.3393\,(34)&   -1.3365\,(55)
   \\\hline\\[-7pt]
 $g^{(0)}_{{\rm vpscr}}$           &              &  0.2628      &  0.2330      &  0.2780      &  0.2104 \\
 $g^{(1)}_{{\rm vpscr}}$           &  0.2614      & -0.0032      &  0.0273      & -0.0183      &  0.0505 \\
 $g^{(2)}_{{\rm vpscr}}$           & -0.0024\,(1) & -0.0005\,(1) & -0.0012\,(1) & -0.0005\,(1) & -0.0019\,(1) \\
 $g_{\rm vpscr}$                   &  0.2590\,(1) &  0.2590\,(1) &  0.2591\,(1) &  0.2591\,(1) &  0.2591\,(1)
%
%
\end{tabular}
\end{table*}

To the leading order in $\Za$,
the effect of the electron self-energy on the bound-electron $g$ factor
of an $s$ state can be represented as
an interaction of the anomalous magnetic moment (AMM) of the electron with the electric and
magnetic fields in the atom \cite{hegstrom:73}.
The effective Hamiltonian is given by
\begin{align}\label{se11}
H_{\rm amm} = \sum_j \Big[H_1(j)+H_2(j) + H_{3}(j)\Big]\,,
\end{align}
where
\begin{align}\label{se12}
H_1(j) = &\ a_e \mu_B \beta_j \bm{B}\cdot \bm{\Sigma}_j\,, \\
H_2(j) = &\ -i\, \frac{a_e}{2} \beta_j\, \bm{\alpha_j}\cdot \bnabla_j \Big[ V_{\rm nuc}(r_j)+U(r_j)\Big]\,,
   \label{se13} \\
H_3(j) = &\ -a_e \frac{\alpha}{2} \sum_{k \ne j} \Big(\beta_j \bm{\Sigma}_j\cdot \frac{\bm{\alpha}_k\times \bm{r}_{jk}}{r_{jk}^3}
 - i\beta_j \frac{\bm{\alpha}_j\cdot \bm{r}_{jk}}{r_{jk}^3}\Big)
  \nonumber \\ &
  +i\, \frac{a_e}{2} \beta_j\, \bm{\alpha_j}\cdot \bnabla_j U(r_j)\,.
  \label{se14}
\end{align}
Here, $j$ and $k$ numerate electrons in the atom,
$a_e = \alpha/(2\pi) + O(\alpha^2)$ is the AMM of the free electron,
$\bm{B}$ denotes the external magnetic field,
$\mu_B = |e|/(2m)$ is the Bohr magneton,
$\bm{r}_{jk} =
\bm{r}_j-\bm{r}_k$, $ \bm{\Sigma} =  \bm{\sigma} {I}$, and ${I}$ is the two-by-two unity matrix.
We note that the screening-potential additions $\propto\!U$ are introduced in the definitions of operators
$H_2$ and $H_3$. While the net effect of these additions is zero, they make the perturbation expansion
consistent with the expansion in the residual electron-electron interaction in Eq.~(\ref{se9}).
Specifically, only $H_1$ and $H_2$ contribute to $g_{\rm sescr}^{(0)}$, whereas $H_3$ can be
interpreted as the AMM correction to the residual electron-electron interaction $I-U$.

Here, we calculate the corrections induced by the AMM operators (\ref{se12})-(\ref{se14}) within
the many-body perturbation theory (MBPT). In order to compute the AMM corrections, we perturb
the standard MBPT formulas for the two-photon exchange correction to the $g$ factor
$g_{\rm int, \rm mbpt}^{(2)}$
by the Hegstrom's
operators multiplied by some parameter $\lambda$ and then linearize the results with respect to
$\lambda$. Specifically, we introduce the following perturbations in the MBPT formulae
\begin{align}
& V_g \to V_g +  \frac{\lambda}{\mu_B\,\mu_v\,B}\,H_1\,,
 \ \ \
 I \to I + \lambda\,H_3\,,
 \nonumber \\ &
\vare_n \to \vare_n + \lambda \bra{n}H_2\ket{n}\,,
 \ \ \
\ket{n} \to \ket{n} + \lambda \sum_{k \ne n}\frac{\ket{k}\bra{k}H_2\ket{n}}{\vare_n-\vare_k}\,.
\end{align}
After performing numerical calculations with different values of $\lambda$,
we linearize the results, obtaining the second-order AMM corrections as
\begin{align}
g_{{\rm sescr}, L}^{(2)} = \frac{\partial g_{\rm int, \rm mbpt}^{(2)}(\lambda)}{\partial \lambda}
 \Big|_{\lambda = 0}\,.
\end{align}

If we restrict our consideration of the second-order screening effects to the lowest order
in $\Za$ only, a significant deviation is observed between the results obtained with the
Coulomb starting potential and screening potentials \cite{kosheleva:22}. This points to large contributions from higher-order (in $\Za$) terms for the Coulomb
starting potential. 
We here calculate a subset of higher-order two-photon screening correction for the
Coulomb potential, specifically, those originating from the one-electron self-energy correction with two and
more interactions with the charge density of the core electrons. These interactions can be accounted for to all orders by a suitably choice of screening
potential. In order to extract the
contribution of two- and more interactions with this potential, we multiply the potential by
a variable parameter $\lambda$, repeat calculations for several values of $\lambda$, and numerically
remove contributions of order $\lambda^0$ and $\lambda^1$. In addition, we calculate this
contribution within the AMM approximation and subtract it, in order to avoid double counting
with $g^{(2)}_{{\rm sescr}, L}$. So, denoting by $g_{\rm se}(\lambda)$ the one-electron self-energy
correction with the core-electron charge density potential multiplied by $\lambda$, we obtain
the higher-order two-photon screening contribution for the Coulomb potential as
\begin{align}
g_{{\rm sescr}, H}^{(2)} =& \
        g_{\rm se}(1) - \frac{g_{\rm se}(\epsilon)+g_{\rm se}(-\epsilon)}{2}
\nonumber \\ &
        - \frac{g_{\rm se}(\epsilon)-g_{\rm se}(-\epsilon)}{2\epsilon}
        - \ldots \mbox{\rm AMM part} \ldots\,,
\end{align}
where $\epsilon$ is a small parameter (typically, $\sim$$10^{-3}$) and the AMM part means
the same expression with the self-energy operator represented by the AMM operators $H_1$ and $H_2$.

Results of our calculations of the self-energy and vacuum-polarization screening corrections
are summarized in Table~\ref{tab:scr}. It presents numerical values of the zero-,
first-, and second-order screening corrections calculated with different starting potentials.
The zero- and first-order corrections are calculated rigorously within QED; their uncertainties
are the estimation of numerical errors. We calculate the second-order correction to the leading
order in $\Za$ for the screening potentials. For the Coulomb potential, this includes
the higher-order (in $\Za$) effects
evaluated as above. The uncertainties of the second-order correction include our
estimations of higher-order effects.

Table~\ref{tab:scr} shows
that the results obtained with different potentials are consistent within the estimated
uncertainties. As final values we select the results
obtained with the core-Hartree (CH) potential. 
We consider this as an optimal starting approximation
for describing the
self-energy and vacuum-polarization screening 
effects since a subset of Feynman diagrams vanish identically with
this choice of the starting potential. This choice is also supported by the observation that the
two-photon screening contribution is the smallest in the CH potential.

%
\subsection{Electron-structure effects}

Electron-structure corrections to the $g$ factor of few-electron atoms are accounted for by QED perturbation theory in the residual
electron interaction $I(\omega) - U$, where the potential $U$ is the screening potential, 
$U = V(r) - V_{\rm nucl}(r)$, i.e.,
the difference of the potential in the Dirac equation $V(r)$ and the nuclear Coulomb
potential $V_{\rm nucl}(r)$.
The QED perturbation expansion of the electron-structure (``int'') effects reads
\begin{align}\label{eq:11}
g_{\rm int} = g^{(0)}_{\rm int} + g^{(1)}_{\rm int} + g^{(2)}_{\rm int} + g^{(3+)}_{\rm int}\,,
\end{align}
where the upper index denotes the order of perturbation expansion (i.e., the number of virtual
photons exchanged between the electrons). The zeroth-order term $g^{(0)}_{\rm int}$ is given by
the difference of the Dirac value $g_{\rm D}$
calculated for the actual starting potential and the Coulomb potential.

The first-order correction $g^{(1)}_{\rm int}$ is induced by the one-photon exchange between
the valence and the core electrons. It is given by \cite{shabaev_g_2002}
\begin{align}  \label{eq:12}
  g^{(1)}_{\rm int}  = & \, 2 \sum_{c}
  \Big[
    \bra{vc}I(0)\ket{\delta{v}c} - \bra{cv}I(\Delta_{vc})\ket{\delta{v}c}
   \nonumber \\ &
    + \bra{vc}I(0)\ket{v\delta{c}} - \bra{cv}I(\Delta_{vc})\ket{v\delta{c}}
   \nonumber \\ &
  - \frac12 \, \bra{cv}I'(\Delta_{vc})\ket{vc}
  \big(\bra{v}V_{g}\ket{v} - \bra{c}V_{g}\ket{c} \big)
  \Big]
   \nonumber \\ &
  -2\,\bra{v} U \ket{\delta v}
  \,,
\end{align}
where $\Delta_{vc} = \varepsilon_v -\varepsilon_c$ is the difference between the Dirac energies of the
valence and core electrons, the prime on $I'(\omega)$ denotes the derivative with respect to the
energy argument, and $\ket{\delta a}$ is the first-order perturbation of the Dirac wave function $\ket{a}$ by the
potential $V_g$.
The one-photon exchange correction was calculated for lithium-like ions in Ref.~\cite{shabaev_g_2002}
and recalculated in Ref.~\cite{cakir:20}. Calculations of this correction are well
established at present and do not pose any difficulties.

The second-order correction $g^{(2)}_{\rm int}$ is induced by an exchange of two photons between
the valence and the core electrons. A rigorous QED calculation of this contribution 
was tackled only relatively recently. Refs.~\cite{volotka:14,kosheleva:22} reported calculations
of this effect for the general screening potential.
In Ref.~\cite{yerokhin:21:gfact} a similar calculation was performed for the Coulomb starting potential
and a wide range of nuclear charge numbers. Formulae for the two-photon exchange correction
were presented in Refs.~\cite{volotka:14,yerokhin:21:gfact} (see also Ref.~\cite{kosheleva:20}).
We here perform an independent calculation of the two-photon exchange correction $g^{(2)}_{\rm int}$
for different starting potentials, in order to independently verify previous results
and improve the numerical accuracy.

\begin{table*}[t]
\smaller
\centering
\caption{
Electron-structure contributions for the $g$-factor of lithium-like tin, for different starting potentials,
in units of $10^{-6}$.\\
\label{tab:elcorr}
}
\begin{tabular}{lw{3.10}w{3.10}w{3.10}w{3.10}w{3.10}}
     \multicolumn{1}{c}{Term} &
         \multicolumn{1}{c}{Coulomb} &
             \multicolumn{1}{c}{CH} &
                 \multicolumn{1}{c}{KS} &
                     \multicolumn{1}{c}{DH} &
                         \multicolumn{1}{c}{DS} \\
                         \hline\\[-5pt]
 $g^{(0)}_{\rm int}$           &                 & 1341.2132      & 1304.7369      & 1370.3162      & 1245.0904       \\
 $g^{(1)}_{\rm int}$           & 1199.6607       & -149.1669      & -110.8615      & -179.4283      &  -49.6789       \\
 $g^{(2)}_{\rm int}$           &   -7.5325\,(4)  &    0.1448\,(4) &   -1.7046\,(4) &    1.3494\,(4) &   -3.2601\,(4)  \\
 $g^{(3)}_{{\rm int}}$         &    0.0296\,(282)&   -0.0115\,(74)&    0.0057\,(74)&   -0.0535\,(99)&    0.0131\,(216)\\
 $g_{\rm int}$                 & 1192.1577\,(282)& 1192.1795\,(74)& 1192.1764\,(74)& 1192.1839\,(99)& 1192.1645\,(216)
\end{tabular}
\end{table*}

The higher-order correction $g^{(3+)}_{\rm int}$ is induced by an exchange by three or more photons
between the valence and the core electrons. These contributions cannot be calculated
rigorously within QED at present.
In Ref.~\cite{kosheleva:22}, this correction was computed within the Breit approximation
by the recursive perturbation theory approach, see Ref.~\cite{glazov:17} for details.
A different method was used in Ref.~\cite{yerokhin:21:gfact}, where it
was calculated for the Coulomb potential strictly to the leading order in $\alpha$
within the framework of the NRQED. Both
methods provide results correct in the leading order in $\alpha$, but the Breit-approximation
calculations also contain admixtures of higher-order $\alpha$ terms.

In the present work, we calculate the correction due to exchange of three and more photons to leading order in
$\alpha$ for a general starting potential. The procedure is as follows. First, for a given starting potential,
we calculate the first three terms of
the perturbative expansion strictly to the leading order in $\alpha$, eliminating all higher-order contributions.
This is achieved by evaluating the non-relativistic limit of the
MBPT corrections. Specifically, we perform
a series of MBPT calculations with different values of the fine-structure constant $\alpha' = x\,\alpha$,
where the prefactor $x$ is varied typically from $x = 0.5$ to $1.5$. After that we fit the results to the
polynomial in $x$ and obtain the non-relativistic limit as the leading contribution $\propto\!x^2$. 
At the second step, we subtract the sum of these perturbation-expansion terms from the complete $\alpha^2$
results evaluated in Ref.~\cite{yerokhin:17:gfact} within the NRQED approach. The difference
gives us the contribution of
three and more photon exchanges for the particular starting potential. We checked that for the Coulomb
potential this procedure reproduces the $1/Z$-expansion of the NRQED results with a very high accuracy.

Our numerical results are summarized in
Table~\ref{tab:elcorr}. The calculations were performed for the Coulomb starting potential as well for the
core-Hartree (CH), Kohn-Sham (KS), Dirac-Hartree (DH), and Dirac-Slater (DS) screening potentials.
In order to provide the uncertainty of the electron-structure correction, we 
estimate the omitted higher-order effects. To be on the safe side, 
we set up two estimations and chose the largest
of them. The first estimate follows Ref.~\cite{kosheleva:22} and is defined by the
magnitude of the higher-order effects for the two-photon exchange correction, scaled by a factor of $2/Z$.
Specifically,
\begin{align}
\delta_{1} g = \frac{2}{Z}\, \big|g_{\rm int}^{(2)} - g_{{\rm int}, L}^{(2)}\big|\,,
\end{align}
where $g_{{\rm int}, L}^{(2)}$ is the two-photon exchange correction evaluated to the lowest-order in $\alpha$.
For the second estimate $\delta_2g$, we take the maximal deviation between
the total electron-structure values $g_{\rm int}$ obtained with the three ``best'' starting
potentials. 
The final estimate of uncertainty is the largest of the two,
$\delta g = \max(\delta_1g,\delta_2g)$.

As the final value of the electron-structure correction we chose the result obtained with the CH potential,
in order to be consistent with the self-energy contribution.
We note that
the numerical uncertainties in the present calculation are well under control so that the
uncertainty of the final value is determined by the estimation of
uncalculated higher-order effects.
%
\subsection{Other effects}

The nuclear recoil correction for lithium-like ions was calculated to all orders in 
$\Za$ in Ref.~\cite{shabaev:17:prl}. The correction 
for the $(1s)^22s$ state of a lithium-like ion is represented as
\begin{align}
g_{\rm rec} = &\, \frac{m}{M}
 \bigg[ 
 \frac23\Big(1+\frac{\vare_v}{m} -2\frac{\vare_v^2}{m^2}\Big) + \frac{(\Za)^5}{8}\,P(\Za)
  + \frac{(\Za)^2}{Z}\,B(\Za)
 + \frac{(\Za)^2}{Z^2}\,C(\Za)\bigg]\,,
\end{align}
where $\vare_v$ is the Dirac energy of the
valence $2s$ electron
and numerical results for the functions $P(\Za)$, $B(\Za)$, and $C(\Za)$
are summarized in Ref.~\cite{shabaev:17:prl}.

The two-loop QED correction is presently known only within the $\Za$ expansion.
The one-electron two-loop QED contribution for the $2s$ state is
\begin{align}
g_{\rm qed2} = &\, \frac{\alpha^2}{\pi^2}\,2\,A^{(2)}
\, \bigg[1 + \frac{(\Za)^2}{24}\bigg]
 + \frac{\alpha^2}{\pi^2} \frac{(\Za)^4}{8}\bigg[\frac{28}{9} \ln (\Za)^{-2}
 \nonumber \\ &
 + c_{40} + (\Za)  \ln (\Za)^{-2} c_{51} + (\Za)\, c_{50} 
 \bigg]\,,
\end{align}
where $A^{(2)} = -0.328\,478\,965\ldots$ is the two-loop contribution to the free-electron $g$ factor,
the $(\Za)^4$ contribution is \cite{pachucki:05:gfact,czarnecki:16}
\begin{align}
c_{40} = -17.157\,236\,58 + \frac{16-19\pi^2}{108}\,,
\end{align}
the logarithmic $(\Za)^5$ contribution is $c_{51} = 28\pi/135$ \cite{czarnecki:20}
and the $(\Za)^5$ contribution is $c_{50} = 12.94\pm1.29$ 
\cite{czarnecki:18,yerokhin:13:twoloopg}. We here ascribe a 10\% uncertainty to $c_{50}$
in order to account for the missing contributions from light-by-light scattering in it, which corresponds to an absolute error of $6\times 10^{-9}$ in the $g$-factor value.
Missing effects of order $\alpha^2(\Za)^{6+}$ are estimated on the basis of known
results for the one-loop QED correction. Specifically, we take the
one-loop contribution of order $\alpha(\Za)^{6+}$, scale it by the ratio
of the two-loop and one-loop $(\Za)^5$ contributions, and multiply the result by a conservative
factor of 2. The resulting error of $\pm 33\times 10^{-9}$ is the dominant
theoretical uncertainty for the $g$ factor of lithium-like tin. 

As an alternative, we also use the experimental value of the $1s$ $g$-factor in hydrogen-like
tin obtained recently in Ref.~\cite{morgner_stringent_2023} in order to obtain the
``experimental'' higher-order two-loop QED contribution and to improve our present
theoretical prediction for the two-loop QED effects in lithium-like tin. Specifically,
by taking the difference of the experimental and theoretical $g$ factor values
from Ref.~\cite{yerokhin:13:twoloopg}, we obtain $0.238\,(26)\times 10^{-6}$ for the
$1s$ two-loop QED contribution of order $\alpha^2(\Za)^{6+}$. In order to
obtain the corresponding contribution for the $2s$ state needed in this work, we
use the $1s$-to-$2s$ ratio of the one-loop corrections of order 
$\alpha(\Za)^{6+}$ for hydrogen-like tin, which is $8.88$.
We thus obtain the $2s$ two-loop QED contribution of order $\alpha^2(\Za)^{6+}$
as $0.238\times 10^{-6} /8.88 = 0.0268\,(61)\times 10^{-6}$,
where the error includes a 20\% uncertainty due to scaling. 
The $1s$-to-$2s$ scaling factor of $8.9\pm 1.8$ is consistent with the typical
$1/n^3$ behavior of QED effects, which originates from their characteristic localization
at small distances close to the nucleus. The 20\% uncertainty estimate of the scaling factor is supported
by the existing calculations of one- and two-loop QED effects to atomic binding energies~\cite{yerokhin_lamb_2015}.

The effect of the screening of the two-loop QED correction is calculated to the
leading order in $\alpha$ by using the Hegstrom's operators (\ref{se11})-(\ref{se14}).
We estimate the uncertainty of the screening correction by comparing with the one-loop case.

%
\subsection{Summary}

\begin{table}[t]
    \centering
    \begin{tabularx}{116mm}{l l r}
         \hline\hline
         Effect & Contribution \hspace{17mm} & Value ($\times 10^{-6}$)\phantom{(xx)}\\\hline
         Electronic structure &  One-electron (pnt) & -23181.721\phantom{(xx)}\\
         & One-electron (FNS) &2.040(3)\phantom{x}\\
         & Many-electron & 1192.179(7)\phantom{x}\\
         1-loop QED &  One-electron &23.977(1)\phantom{x}\\
         & Many-electron (SE) & -1.336(2)\\
         & Many-electron (VP) & 0.259\phantom{(xx)}\\
         2-loop QED & One-electron & -0.107(33)\\
         \,\,\,-\,\,\, , enhanced &  & -0.080(6)\phantom{x}\\
         & Many-electron & 0.001(8)\phantom{x}\\
         $\geq$3-loop QED & One-electron & \phantom{-}0.000(1)\phantom{x}\\
         Recoil & One-electron & 0.178\phantom{(xx)}\\
         & Many-electron & -0.006(1)\phantom{x}\\ \hline
         Total & $g$ & 1980354.769(35)\\
         & $g$, enhanced & 1980354.797(12)\\ 
         & $g_\mathrm{exp}$ & 1980354.800(1)\phantom{x}\\ \hline\hline
    \end{tabularx}
    \caption{
        Binding corrections to the $g$ factor of lithium-like tin.
        The sum of all contributions is the difference between the atomic $g$ factor and the free-electron $g$ factor, $g_e = 2.002\,319\,304\,361\,2(3)$.
        }
    \label{tab:gtotal}
\end{table}
In Table~1 we collect all known binding corrections to the $g$-factor
of the ground state of lithium-like tin. Various effects
are separated into
the one-electron and many-electron parts. The one-electron parts are calculated
for the valence $2s$ hydrogenic state, 
with the nuclear Coulomb potential that includes the finite nuclear size effect. 
The many-electron effects are induced by the presence of the core electrons and
are calculated as described above.

Results for the one-electron two-loop QED effects are presented in Table~1
in two variants. The first value
of $-0.107\,(33)\times 10^{-9}$ is the purely theoretical result, whereas
the second value $-0.080\,(6)\times 10^{-9}$ is obtained with the use of the
experimental result for the $1s$ $g$-factor \cite{yerokhin:13:twoloopg}. We observe that
the two values are in good agreement with each other but the ``experimentally 
enhanced'' result is 5.5 times more accurate.

\section{Summary of the experiment}
Overall, the techniques and methods used here are identical to those in Ref.~\cite{morgner_stringent_2023}.
The axial frequency, set to the corresponding detection systems, is around 652\,kHz in the Precision Trap, and 334\,kHz in the Analysis Trap.
The radial modes in the Precision Trap are around 8.6\,kHz for the magnetron, and 24.6\,MHz for the modified cyclotron frequency.
The measurement of the magnetron motion is done only every few cycles, therefore, we conservatively assume a large uncertainty on the value, resulting in the 4.3\,ppt uncertainty given in Table.~2 of the main text.\\
The mentioned temperature of 5.5(2)\,K results from the continuous probing of the cyclotron energy in the Analysis Trap.
A similar shift as in Eq.~\ref{eq:deltawz} arises also from the magnetic moment of the cyclotron orbit, and a different radius results therefore in a shifted axial frequency:
\begin{equation}\label{eq:deltawz2}
    \Delta \nu_z \approx \frac {B_2\hbar q}{4 \pi^2 M^2 \nu_z}\hat n_+,
\end{equation}
with $\hat n_+$ being the modified cyclotron quantum number.
From the Boltzmann distribution of the cyclotron mode after sideband cooling via the axial mode one can infer its temperature.
From the coupling, the axial energy is then given as $T_z \approx \nicefrac{\nu_\mathrm z}{\nu_+}T_+$~\cite{ketter_first-order_2014}.\\
The next important uncertainty in the $g$-factor measurement arises from the axial frequency measurement, which is measured before and after the $\nu_+$ determination.
In both, the axial frequency is measured by fitting the Fourier noise spectrum of the cryogenic detector.
This 'dip'-fitting is prone to systematic shifts, as in example a frequency-dependent transfer function would cause a systematic shift of the measured dip position.
Following Ref.~\cite{morgner_stringent_2023}, we give a 20\,mHz uncertainty on the axial frequency, resulting in a 22\,ppt uncertainty on $\Gamma_0$.\\
Two other non-negligible systematic effects are on the one hand the image-charge shift, causing a shift of the motions due to the image-charges induced in the surrounding electrodes~\cite{schuh_image_2019}.
This is corrected according to the theoretical prediction of the shift, leaving an uncertainty of 7.5\,ppt on the final result.
The other is the relativistic effect that has to be taken into account~\cite{ketter_classical_2014}.
During the $\nu_\mathrm p$ measurement, the radius is increased to about 12.8\,µm, which we give a conservative error of 10\,\%, resulting in a higher kinetic energy, and hence a relativistic contribution
\begin{equation}
    \frac{\Delta \nu_\mathrm c}{\nu_\mathrm c} \approx -\frac{E_+}{Mc^2},
\end{equation}
with $E_+$ as the energy of the modified cyclotron mode.
This is corrected, leaving a residual uncertainty of 4.4\,ppt.
Similarly, effects from the thermal radius cause a relativistic shift but are small compared to the statistical uncertainty.\\
Electric imperfections have little effect on the result due to the large trap, and the small thermal radius of the particle due to its weight.
From a dedicated measurement, we infer the anharmonicity coefficient (for a broader description see e.g. Ref.~\cite{ketter_first-order_2014}) $C_4$ to be smaller than $2.5\times 10^{-5}$ with the characteristic trap length being set to the trap radius of 9\,mm.
Similar $C_6$ is measured to be less than $2\times 10^{-3}$.
Asymmetric terms, most dominantly $C_3$, have been studied in the trap as well providing no limit on the presented accuracy~\cite{sailer_measurement_2022,sailer_direct_2022}.
With the temperature as described above, the resulting shift on the frequencies -- as given in Ref.~\cite{ketter_first-order_2014} -- changes the final result by less than 1\,ppt;
The second-order magnetic-field inhomogeneity is measured to be less than 10\,$\mathrm{mT}/\mathrm{m}^2$, and equally has a negligible effect.\\
Following the discussion in Ref.~\cite{morgner_stringent_2023}, the line width of the measured resonance is dominated by the magnetic field jitter.
This presents itself as a Gaussian distribution in the line shape.
The microwave power is set to cause less than 50\,\% spin-flip probability in the center of the resonance.
The underlying Lorentzian shape originating from driving the transition is therefore suppressed and the Gaussian dominates, allowing to fit the resonance probability curve $P(\Gamma)$ with a pure Gaussian profile
\begin{equation}
    P(\Gamma) = A \mathrm e^{-\frac{(\Gamma-\Gamma_\mathrm{stat})^2}{2\sigma^2}}.
\end{equation}
Here, $A$ is the amplitude, fitted to 31(6)\,\%.
The extracted full-width half-maximum $2\sqrt{2 \mathrm{ln}(2)}\sigma$ is 538(32)\,ppt.
With the statistical uncertainty on the center of $4.2\times10^{-11}$ the line splitting is about a factor of 13.
Line width contributions from the magnetic field inhomogeneity are so small that their contribution does not effect the final result.

\end{document}